\documentclass[reprint,twocolumn,amsmath,showpacs,amssymb]{revtex4-1}
\usepackage{graphicx,bm}
\usepackage[hidelinks]{hyperref}

\def\unitkelvin{\,\textrm{K}}
\def\unittep{\,{\rm \mu V/K}}
\def\unitev{\,{\rm eV}}

\def\unitnm{\,{\rm nm}}

\def\mod{\textrm{mod}}
\def\diameter{d_{\rm t}}

\begin{document}

\title{Diameter dependence of thermoelectric power of semiconducting
  carbon nanotubes}
 
\author{Nguyen T. Hung$^1$}
\email{nguyen@flex.phys.tohoku.ac.jp}

\author{Ahmad R. T. Nugraha$^1$}
\email{nugraha@flex.phys.tohoku.ac.jp}

\author{Eddwi H. Hasdeo$^1$}
\author{Mildred S. Dresselhaus$^{2,3}$}
\author{Riichiro Saito$^1$}

\affiliation{$^1$Department of Physics, Tohoku University, Sendai
  980-8578, Japan\\ $^2$Department of Electrical Engineering,
  Massachusetts Institute of Technology, Cambridge, MA 02139-4307,
  USA\\ $^3$Department of Physics, Massachusetts Institute of
  Technology, Cambridge, MA 02139-4307, USA}

\begin{abstract}
  We calculate the thermoelectric power (or thermopower) of many
  semiconducting single wall carbon nanotubes (s-SWNTs) within a
  diameter range $0.5$--$1.5\unitnm$ by using the Boltzmann transport
  formalism combined with an extended tight-binding model. We find
  that the thermopower of s-SWNTs increases as the tube diameter
  decreases. For some s-SWNTs with diameters less than $0.6\unitnm$,
  the thermopower can reach a value larger than $2000\unittep$ at room
  temperature, which is about $6$ to $10$ times larger than that found
  in commonly used thermoelectric materials.  The large thermopower
  values may be attributed to the one-dimensionality of the nanotubes
  and to the presence of large band gaps of the small-diameter
  s-SWNTs.  We derive an analytical formula to reproduce the numerical
  calculation of the thermopower and we find that the thermopower of a
  given s-SWNT is directly related with its band gap.  The formula
  also explains the shape of the thermopower as a function of tube
  diameter, which looks similar to the shape of the so-called Kataura
  plot of the band gap dependence on tube diameter.
\end{abstract}

\pacs{79.10.-n,72.20.Pa,65.80.-g}
\date{\today}
\maketitle

\section{Introduction}
In recent years, there has been significant interest in research on
thermoelectric phenomena due to the increase in the demand for
alternative energy sources.  Especially, since thermoelectric
phenomena could transform heat currents into electric power,
thermoelectric power generators can perhaps be used to convert waste
heat into electric energy for use in environmentally friendly
applications~\cite{heremans13-thermo,vining09-thermo,majumdar04-thermo}.
It is thus necessary to find a good thermoelectric material with a
high thermoelectic energy conversion efficiency, characterized by the
so-called thermoelectric figure of merit, $ZT = S^2 \sigma \kappa^{-1}
T$, where $S$ is the Seebeck coefficient, also known as the
thermoelectric power (thermopower), $\sigma$ is the electrical
conductivity, $\kappa$ is the thermal conductivity, and $T$ is the
absolute temperature of the material. Over the past six decades it has
been challenging to obtain $ZT$ values exceeding 2, because the
parameters of $ZT$ are generally
interdependent~\cite{vining09-thermo,majumdar04-thermo}. A theoretical
study in 1993 predicted that the $ZT$ value of low-dimensional
structures could be significantly enhanced, thanks to the quantum
confinement effect to create sharp features in the density-of-states
(DOS)~\cite{hicks93-thermo}. This prediction was confirmed
experimentally in 1996 using PbTe/Pb$_{1-x}$Eu$_{x}$Te, which
exhibited a $ZT$ value up to about five times greater than that of the
corresponding bulk value~\cite{hicks96-thermo}. It is thus intriguing
to evaluate other low-dimensional structures that might have excellent
thermoelectric performance, either theoretically or experimentally.

As a one-dimensional material, single wall carbon nanotubes (SWNTs)
were considered promising for thermoelectric materials due to their
novel electronic properties which depend on their geometrical
structure~\cite{hone98-thermobulk,hone00-thermoapl,yanagi14-thermofilm}.
However, it has been difficult to obtain an ensemble of individual
SWNTs with a specific $(n,m)$ structure to reveal the precise
knowledge of the dependence of the thermoelectric power of individual
SWNTs on band gap and diameter.  Most thermoelectric measurements were
performed on bundled SWNT samples whose geometrical and electronic
structures are
complex~\cite{hone98-thermobulk,hone00-thermoapl,yanagi14-thermofilm,romero02-thermofilm},
and thus the potential thermoelectric properties might have been lost
because of interactions between different
tubes~\cite{hone98-thermobulk}. The $ZT$ values reported for bundled
SWNTs have remained in the range of $10^{-3}$ to
$10^{-4}$~\cite{romero02-thermofilm,zhang07-thermobulk}, in contrast
to the commercial thermoelectric materials with $ZT \approx
1$~\cite{dresselhaus07-thermo,poudel08-thermo}.  Such bundled SWNT
samples consist of a collection of SWNTs with different diameters,
metallicities, and chiralities, parameters to which the electronic
structure is very sensitive~\cite{saito98-phys}. The small $ZT$ value
of the bundled SWNTs were mainly attributed to their low thermopower
and high thermal conductivity, which might be a result of the mixture
of different SWNTs and impurity in low concentration in the samples.

In this work, we will focus on evaluating the thermopower
theoretically for many SWNTs, especially in the case of semiconducting
SWNTs (s-SWNTs), and thus to maximize the SWNT thermopower and to
suggest a new route for obtaining a larger $ZT$ for SWNTs.  By
calculating the thermopower of all individual s-SWNTs within a
diameter range $0.5\le\diameter\le1.5\unitnm$, we will show that, for
tube diameters less than $0.6\unitnm$ under low doping, the
thermopower of s-SWNTs can be as large as $2000\unittep$ at room
temperature, which is large enough compared to the thermopower of
bundled SWNTs, which is about
$100-200\unittep$~\cite{hone98-thermobulk,yanagi14-thermofilm,nakai14-giantS}.
From this result, we believe that there is still much room available
to improve the $ZT$ of SWNT samples.  For a more practical purpose, we
also give an analytical formula to reproduce our numerical calculation
of the s-SWNT thermopower, which forms a map of the s-SWNT
thermopower.  The calculated thermopower map could be useful for
obtaining information on the s-SWNT chirality with a desired
thermopower value and thus it offers promise for using specially
prepared s-SWNT samples to guide the direction of future research on
the thermoelectricity.

This paper is organized as follows.  In Sec.~\ref{sec:model}, we give
the theoretical methods employed in this study to calculate the
thermopower.  In Sec.~\ref{sec:results}, we discuss the thermopower
obtained from the numerical calculation and compare it with the
analytical formula.  We then summarize the results and give the future
perspective in Sec.~\ref{sec:summary}.  We also provide some
appendices for a detailed derivation of the thermopower analytical
formula.

\begin{figure}[t]
  \centering
  \includegraphics[clip,width=85mm]{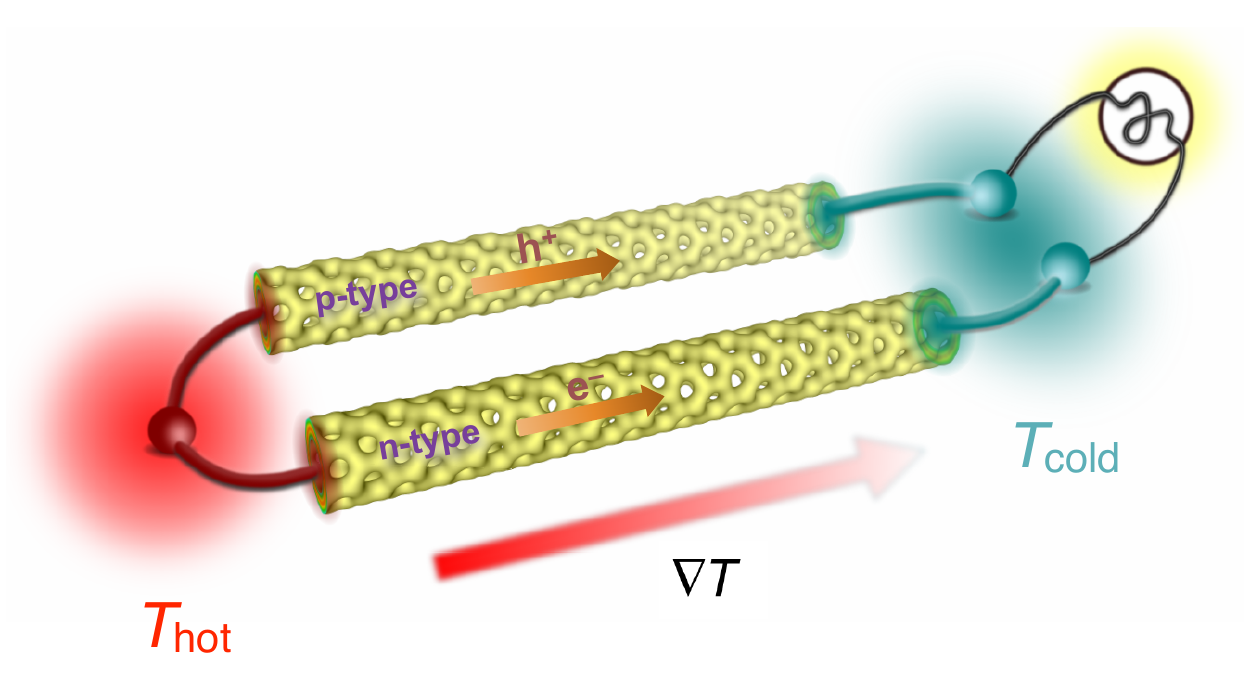}
  \caption{\label{fig:model} (Color online) Schematic model of a
    thermoelectric device using two identical s-SWNTs, one with p-type
    and the other with n-type doping.  The temperature gradient
    between the two edges of each nanotube generates an electric
    current.}
\end{figure}

\section{Model and Methods}
\label{sec:model}

To utilize the s-SWNTs as a main material in future thermoelectric
devices, we consider a model shown in Fig.~\ref{fig:model}, in which
two identical s-SWNTs, one with p-type and the other with n-type
doping, are connected in parallel.  Each s-SWNT should maintain its
electronic charge distribution in the nonequilibrium state, for
example, by a temperature gradient along the tube axis.  By having
their temperature gradient $\nabla T$ from an edge of each s-SWNT to
its other edge, charge carriers (electrons or holes) will flow with
velocity $v$ from the hot edge with temperature $T_{\rm hot}$ to the
cold edge with temperature $T_{\rm cold}$.  The carrier distribution
$f_0$, which depends on the electronic energy $\varepsilon$ and
chemical potential $\mu$, is modified as a function of $\varepsilon$,
following the Boltzmann transport formalism.  Within such a process,
an electric voltage $\nabla V$ can be generated.  It is also known
from earlier studies that the electron-phonon interaction is the main
factor determining the electrical conductivity of
SWNTs~\cite{suzuura02-phonon,javey04-phonon,jiang05-elph}, in which
the so-called twisting (TW) phonon mode with a long wavelength gives
the dominant contribution to the electron-phonon interaction.  In
particular, Jiang \emph{et al.} showed that the relaxation time from
the electron scattering with the TW phonon mode is independent of the
electron energy~\cite{jiang05-elph}.  Therefore, here we make the
assumption that the thermopower from the Boltzmann transport equation
can be obtained by applying the relaxation time approximation (RTA)
and we may even treat the relaxation time as a constant.  Under the
RTA, the thermopower or Seebeck coefficient $S$ is expressed by
\begin{align}
  S &= -\frac{\nabla V}{\nabla T}\notag\\
  &= \frac{1}{q T} \frac{\displaystyle \int v(\varepsilon) \tau
    (\varepsilon) v(\varepsilon) \frac{\partial
      f_0(\varepsilon)}{\partial \varepsilon} [\varepsilon - \mu]
    g(\varepsilon) d\varepsilon}{\displaystyle \int v(\varepsilon)
    \tau(\varepsilon) v(\varepsilon) \frac{\partial
      f_0(\varepsilon)}{\partial \varepsilon} g(\varepsilon)
    d\varepsilon},
\label{eq:seebeck}
\end{align}
where $q = \pm e$ is the unit carrier charge, $T = (T_{\rm hot} +
T_{\rm cold})/2$ is the average absolute temperature, $v(\varepsilon)$
is the carrier velocity, $g(\varepsilon)$ is the electronic (DOS), and
$\tau(\varepsilon)$ is the carrier relaxation time.

We employ both numerical and analytical methods to obtain $S$ from
Eq.~\eqref{eq:seebeck}.  In the full numerical approach, we can use
the BoltzTraP code~\cite{madsen06-boltztrap}, which is a widely-used
package to calculate some thermoelectric properties, such as the
thermopower and electrical conductivity.  A necessary input for the
BoltzTraP code is the electronic energy dispersion $\varepsilon(k)$
for all bands (multiband structure).  The BoltzTraP code also adopts a
constant $\tau$, whose plausibility in the case of s-SWNTs has been
justified above.  While the BoltzTraP code is actually sufficient for
obtaining the thermopower from Eq.~\eqref{eq:seebeck}, we cannot
discuss the physics of the thermopower of s-SWNTs without having an
explicit formula for the thermopower that depends on some physical
parameters, such as the SWNT band gap and geometrical structure.
Therefore, we also solve Eq.~\eqref{eq:seebeck} analytically by
considering the valence band and the conduction band closest to the
Fermi energy, known as the two-band
model~\cite{goldsmid99-thermo,goldsmid10-thermo}.  The derivation of
the analytical formula is explained in detail in Appendices A-D.

As the input for the BoltzTraP code, we calculate the energy
dispersion $\varepsilon (k)$ within the extended-tight binding (ETB)
model developed in our group~\cite{georgii04-etb}.  The ETB model
takes into account long-range interactions, SWNT curvature
corrections, and geometrical structure optimizations, which are
sufficient to reproduce the experimentally observed energy band gaps
of the SWNTs~\cite{georgii04-etb,popov04-etb,weisman03-opt}.  The SWNT
structure in our notation is denoted by a set of integers $(n,m)$
which is a shorthand for the chiral vector $\textbf{C}_h =
n\textbf{a}_1 + m\textbf{a}_2$, where $\textbf{a}_1$ and
$\textbf{a}_2$ are the unit vectors of an unrolled graphene
sheet~\cite{saito98-phys}. The chiral vector $\textbf{C}_h$ defines
the circumferential direction of the tube, giving the diameter
$\diameter$.  Another vector perpendicular to $\textbf{C}_h$ defines
the tube axis, which is called the translational vector
$\textbf{T}$~\cite{saito98-phys}. The chiral and translational vectors
thus represent the tube unit cell.  In the BoltzTraP calculation, we
use a $20\unitnm \times 20 \unitnm \times |\textbf{T}|$ supercell,
where $|\textbf{T}|$ (in nm) is the length of the translational
vector.  A large supercell length in the $x$- and $y$-directions is
chosen so as to guarantee the individual SWNTs are well-separated.  Since
the thermopower in the BoltzTraP code is expressed in terms of a
tensor~\cite{madsen06-boltztrap}, the corresponding thermopower tensor
component for a given s-SWNT is $S_{zz}$, which is the thermopower
along the tube axis direction.  Other tensor components are
negligible.

\begin{figure}[t]
  \centering
  \includegraphics[clip,width=85mm]{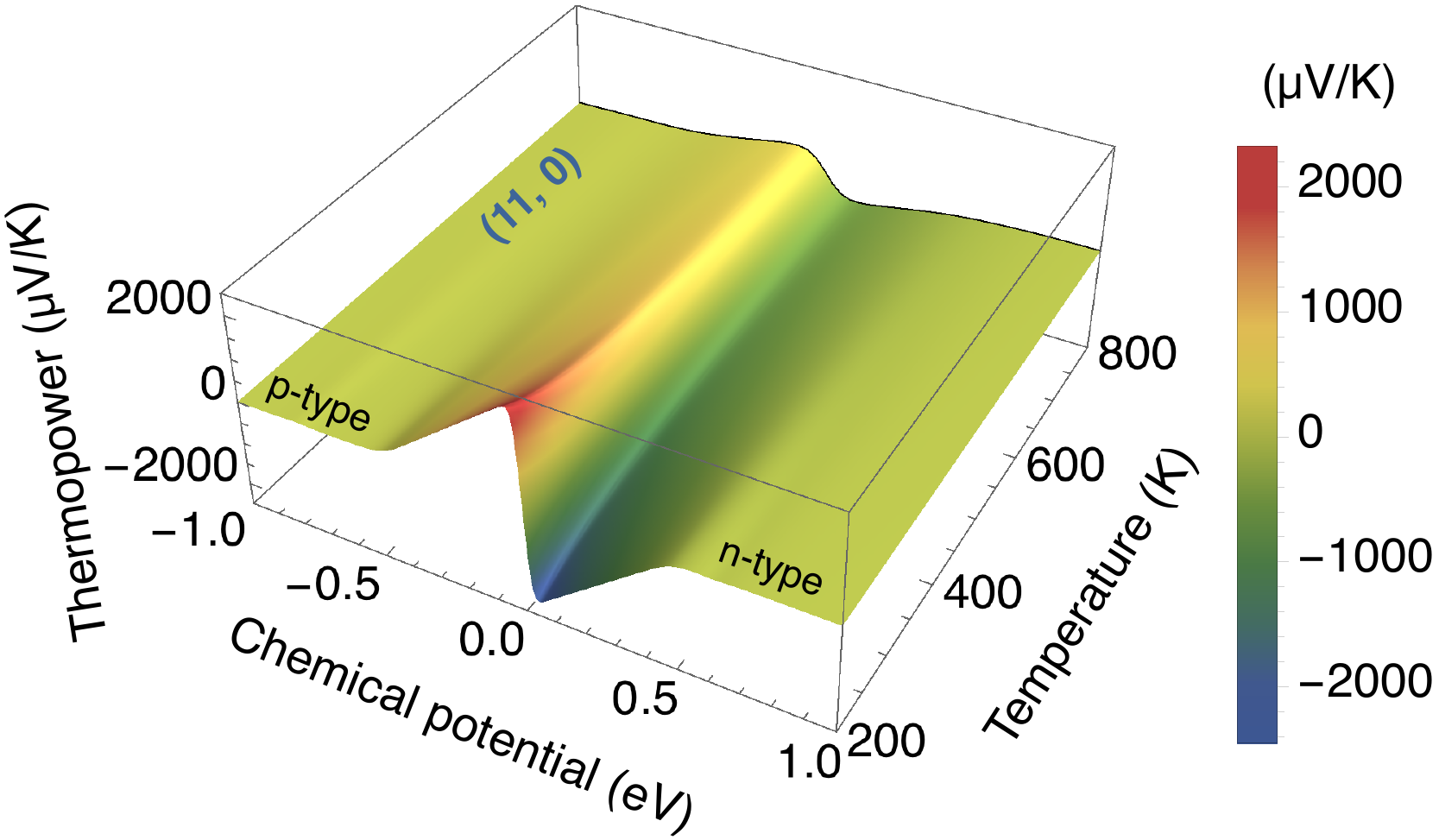}
  \caption{\label{fig:S-T} (Color online) Thermopower as a function of
    chemical potential and temperature for an $(11,0)$ s-SWNT.}
\end{figure}

\section{Results and Discussion}
\label{sec:results}

\begin{figure}[t]
  \centering
  \includegraphics[clip,width=85mm]{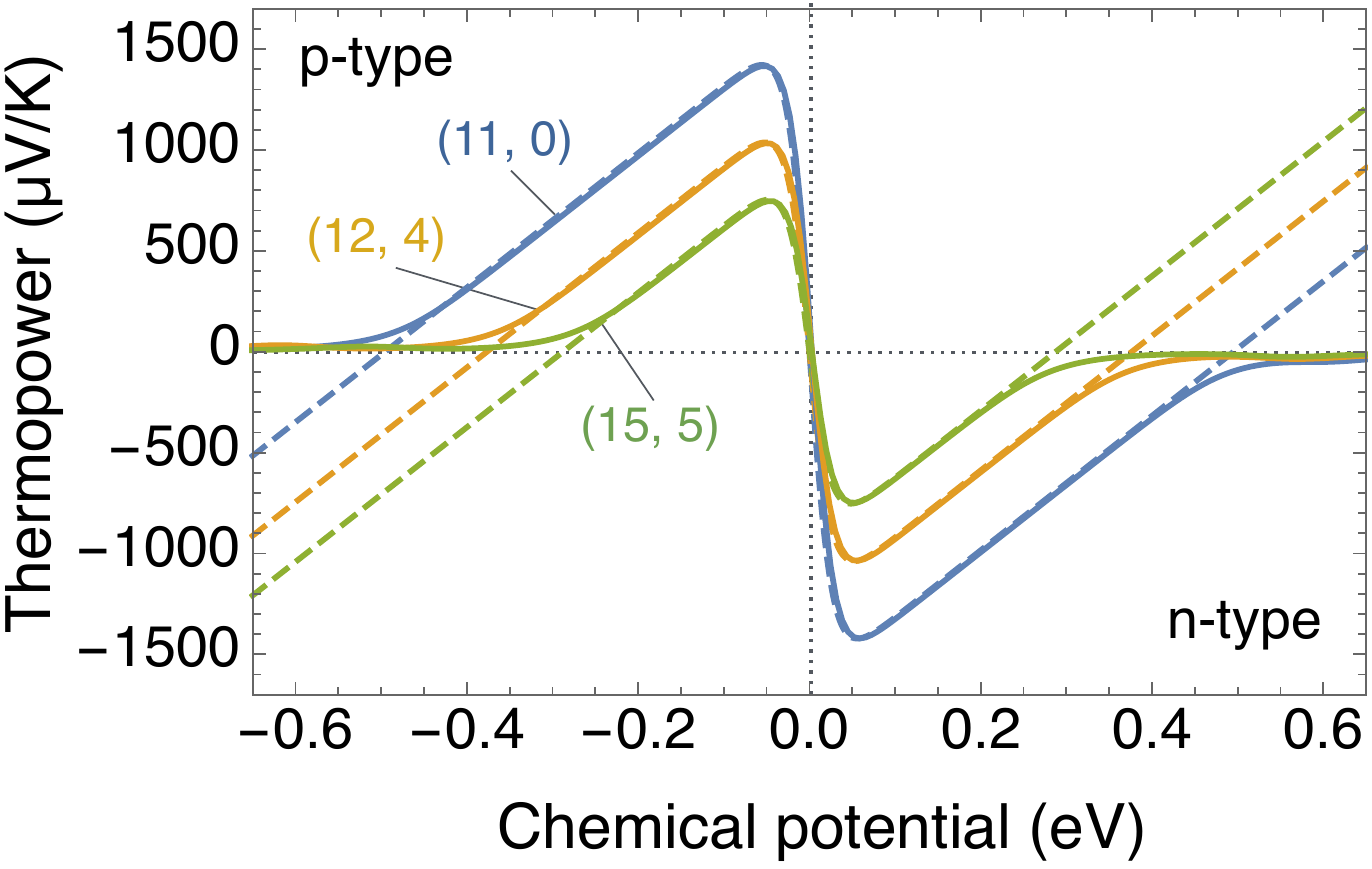}
  \caption{\label{fig:ETB-CNT} (Color online) Thermopower as a
    function of chemical potential for $(11,0)$, $(12,4)$, and
    $(15,5)$ s-SWNTs at $300\unitkelvin$. Solid lines are obtained
    from the numerical calculation based on Eq.~\eqref{eq:seebeck}
    while dashed lines are obtained from the analytical formula given
    in Eq.~\eqref{eq:scnt}.}
\end{figure}

In Fig.~\ref{fig:S-T}, we show a first example of the thermopower
calculation result for an $(11,0)$ s-SWNT.  The thermopower ($S_{zz}$)
is plotted versus chemical potential and temperature.  We see that the
thermopower is higher at the lower temperature because $S \propto 1/T$
in Eq.~\eqref{eq:seebeck}. The maximum thermopower obtained for the
$(11,0)$ SWNT is about $1420\unittep$, which is already large for a
purely individual s-SWNT compared to that for bundled SWNTs with $S$
of around
$100-200\unittep$~\cite{hone98-thermobulk,yanagi14-thermofilm}.  Next,
we can also plot the thermopower at a specific temperature to see the
chemical potential dependence of the thermopower.  In
Fig.~\ref{fig:ETB-CNT}, we show the thermopower versus chemical
potential for three different s-SWNT chiralities: $(11,0)$, $(12,4)$,
and $(15,5)$, at $T = 300\unitkelvin$.  The solid lines in
Fig.~\ref{fig:ETB-CNT} represent the numerical results.  For all
chiralities, the optimum value of the thermopower, indicated by a
maximum (minimum) along the negative (positive) axis of the chemical
potential, arises due to the p-type (n-type) characteristics of the
s-SWNTs, which is consistent with a recent experimental
observation~\cite{yanagi14-thermofilm}. The dependence of the
thermopower on the chemical potential implies that it is possible to
tune the thermoelectric properties of s-SWNTs by applying a gate
voltage, giving p-type and n-type control over the thermopower.

To better understand the numerical results of thermopower, we have
derived an analytical formula for the thermopower within the two-band
model~\cite{goldsmid99-thermo,goldsmid10-thermo}.  We denote this
analytical formula of the thermopower as $S_\mathrm{CNT}$ (see
Appendices A-D for the detailed derivation).  The final form of
$S_\mathrm{CNT}$ can be written as
\begin{align}
  S_\mathrm{CNT} = \frac{k_B}{e} \left(\frac{\mu}{k_B T} -
    \frac{E_g}{2 k_B T} - \frac{3}{2} +\frac{E_g/k_B T +3}{e^{2\mu/k_B
        T} +1}\right),
\label{eq:scnt}
\end{align}
where $e$ is the elementary electric charge, $k_B$ is the Boltzmann
constant, and $E_g$ is the SWNT band gap.  The $E_g$ values adopted in
Eq.~\eqref{eq:scnt} are obtained from previous ETB
results~\cite{georgii04-etb}. The dashed lines in
Fig.~\ref{fig:ETB-CNT} represent the fit of the numerical results of
the thermopower using Eq.~\eqref{eq:scnt} for three different s-SWNT
chiralities.  The analytical formula [Eq.~\eqref{eq:scnt}] fits the
numerical results near $\mu = 0$.  In particular, the two optimum
thermopower values (maximum and minimum for p-type and n-type doping,
respectively) can be well-reproduced in that region, which implies
that the energy bands near the Fermi energy give the strongest
contribution to the thermopower of s-SWNTs. The analytical results
deviate from the numerical results at larger $|\mu|$ far from the
optimum thermopower because the two-band model is no longer valid at a
higher doping level.  However, for the discussion in this paper, the
two-band model is already sufficient to describe the thermopower of
s-SWNTs since we will mainly focus on the optimum values of the
thermopower.

\begin{figure}[t]
 \centering
 \includegraphics[clip,width=85mm]{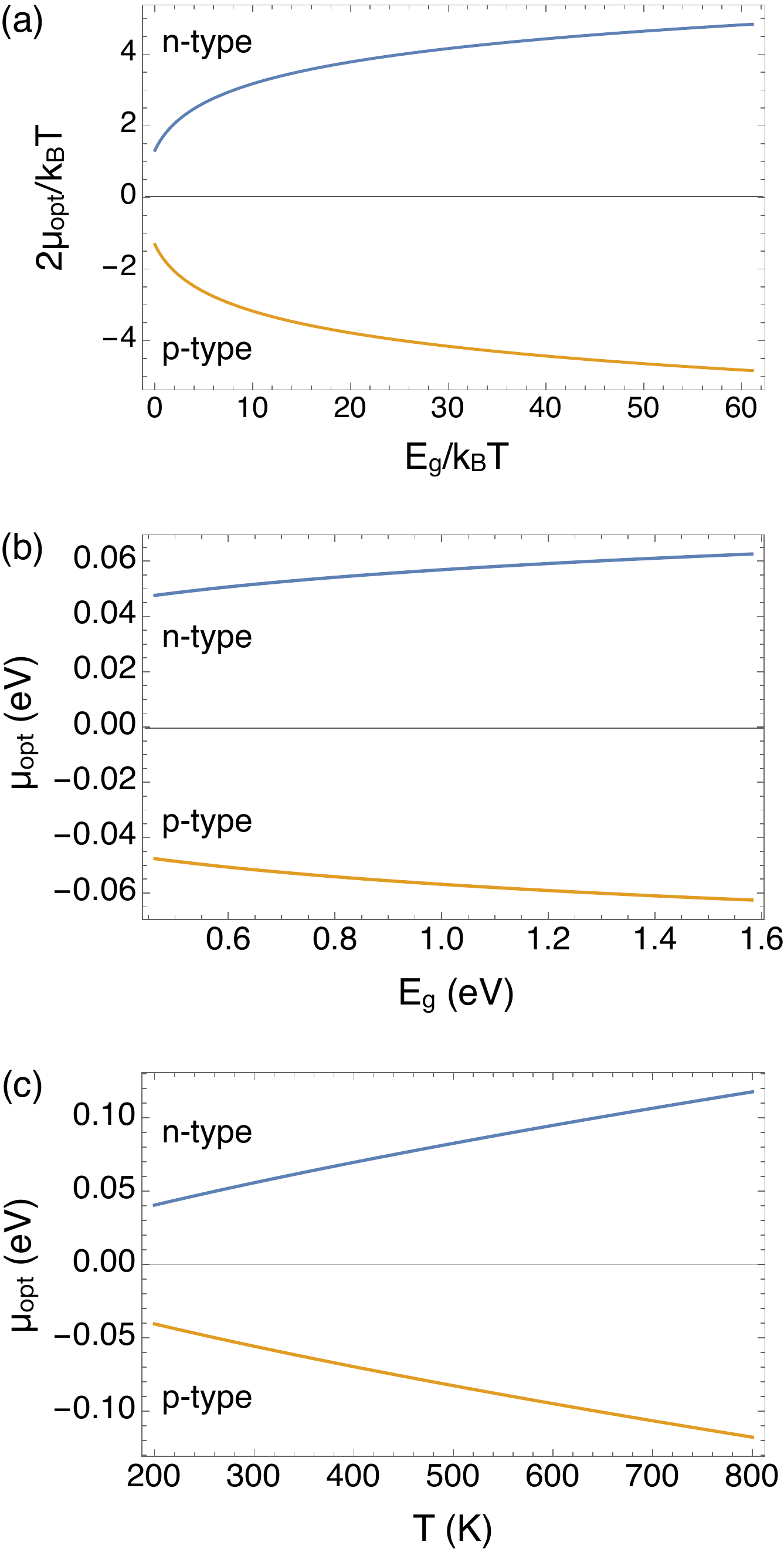}
 \caption{\label{fig:muopt} (Color online) The optimized chemical
   potential $\mu_{\rm opt}$ plotted as a function of s-SWNT band gap.
   In panel (a), we scale the chemical potential and the band gap by
   $k_BT/2$ and $k_B T$, respectively, as described by
   Eq.~\eqref{eq:muopt}.  In the case of (b), we set a constant $T =
   300$ K and vary $E_g$, while in (c) we set a constant $E_g =
   0.913\unitev$, which is the band gap value of an $(11,0)$ s-SWNT,
   and vary the temperature.}
\end{figure}

For a more rigorous argument, we determine a condition to obtain an
optimized chemical potential $\mu_{\rm opt}$ from Eq.~\eqref{eq:scnt},
which satisfies $dS_\mathrm{CNT} (\mu_{\rm opt})/d\mu = 0$.  We then
obtain
\begin{align}
\mu_\mathrm{opt} = \frac{k_B T}{2} \ln \Bigg( \frac{E_g}{k_B T}
+ 2 \pm \sqrt{\Big(\frac{E_g}{k_B T} + 2\Big)^2 -1} \Bigg),
\label{eq:muopt}
\end{align}
where the $+$ and $-$ signs define the n-type and p-type
contributions, respectively.  From Eq.~\eqref{eq:muopt}, we can say
that the $\mu_\mathrm{opt}$ values will move more distant from $\mu =
0$ as $E_g$ becomes larger than $k_B T$, as shown in
Fig.~\ref{fig:muopt}(a).  However, due to the presence of the
logarithmic term, $\mu_\mathrm{opt}$ is very slowly changing as a
function of $E_g$ when $E_g$ is much larger than $k_B T$.  This
behavior can be seen in Fig.~\ref{fig:muopt}(b), in which we show the
$E_g$ dependence of $\mu_\mathrm{opt}$.  For the $d_t$ range of
$0.5$--$1.5\unitnm$, the s-SWNTs have $E_g$ values of about
$1.58\unitev$ down to $0.46\unitev$.  In this case, $E_g$ is about
$17$--$61$ times larger than $k_B T$ for $T = 300$ K.  With those
$E_g$ values, we then obtain $0.046<|\mu_{\rm opt}|<0.062\unitev$ at a
constant $T = 300\unitkelvin$ [see Fig.~\ref{fig:muopt}(b)], which
implies that the change in $\mu_{\rm opt}$ in this case is only about
$16$ meV although the change in $E_g$ is as large as about
$1.12\unitev$ for the same $d_t$ range.  At room temperature,
controlling the doping level or the chemical potential is thus useful
to give us the optimum thermopower for the s-SWNTs under
consideration.  On the other hand, by decreasing $T$ for a given
$E_g$, we can also decrease $\mu_\mathrm{opt}$, as shown in
Fig.~\ref{fig:muopt}(c), which reduces the doping level required to
obtain the optimum thermopower.  It should be noted that in
Fig.~\ref{fig:muopt}(c) we intentionally set a constant $E_g =
0.913\unitev$ for simplicity although the s-SWNT band gaps in the
realistic case may decrease as a function of temperature by about
$3\%$ when we increase $T$ from $200\unitkelvin$ to
$800\unitkelvin$~\cite{capaz05-tempeg}.

\begin{figure}[t]
  \centering
  \includegraphics[clip,width=85mm]{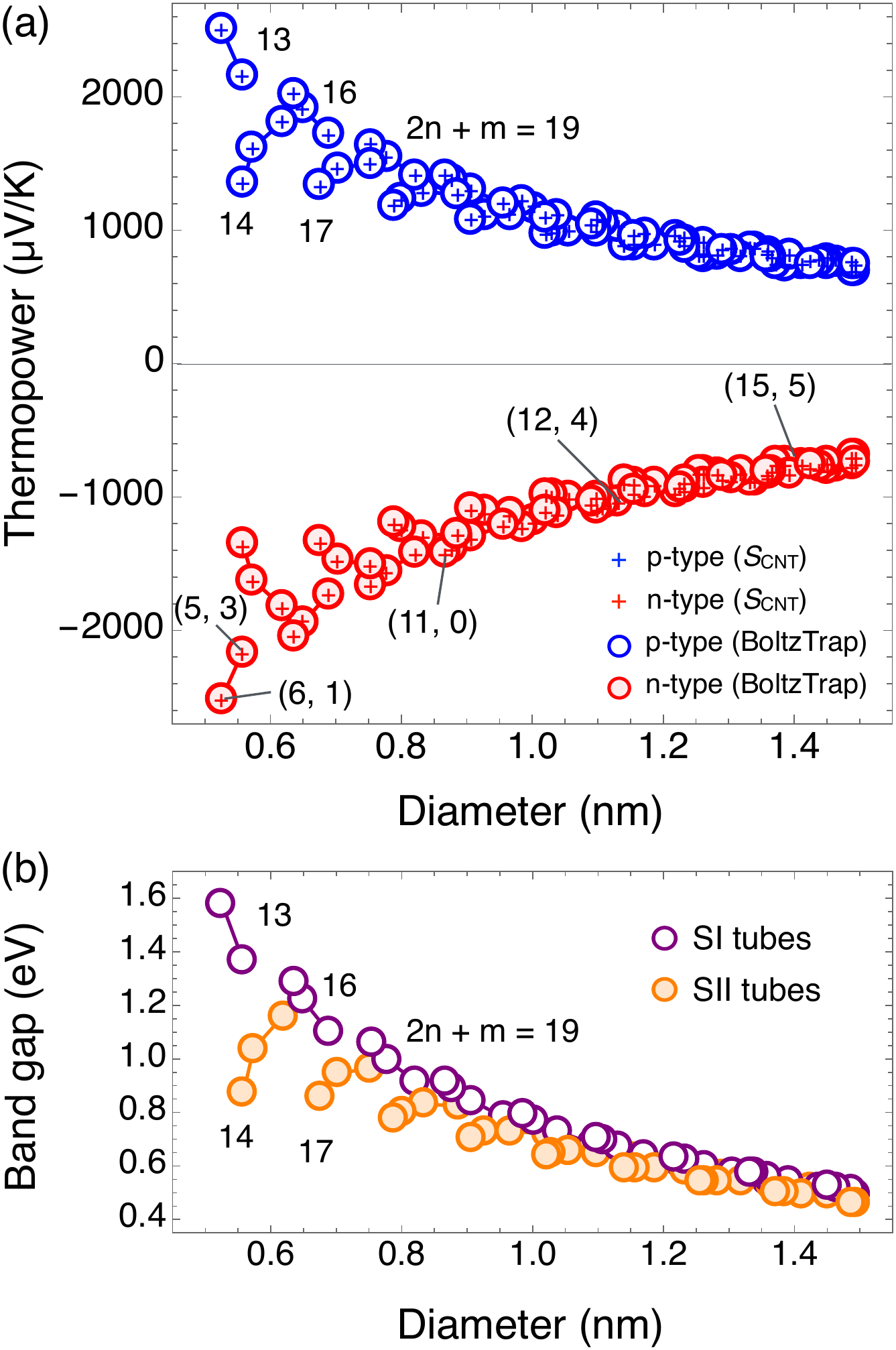}
  \caption{\label{fig:thermo-plot} (Color online) (a) Optimum
    thermopower $S_{\rm CNT}^{\rm opt}$ values for all s-SWNTs within
    the diameter range of $0.5$--$1.5\unitnm$ plotted as a function of
    SWNT diameter.  The temperature is set constant at
    $300\unitkelvin$.  Numerical results from BoltzTraP are denoted by
    circles, while analytical results from
    Eqs.~\eqref{eq:scnt}-\eqref{eq:sopt} are denoted by plus symbols.
    (b) The Kataura plot showing the family pattern of the SWNT band
    gap as a function of diameter.  Solid lines are a guide for the
    eyes, connecting SWNTs with the same family number $2n+m$. The SI
    and SII tubes correspond to the SWNTs having $\mod(2n+m,3) = 1$
    and $2$, respectively.}
\end{figure}

Using both the numerical calculation by BoltzTraP and our analytical
formula $S_\mathrm{CNT}$, it is now possible for us to plot the
thermopower of s-SWNTs over a broad range of $d_t$ by taking the
optimum value of the thermopower.  In the case of the analytical
formula, we define the optimum thermopower $S_{\rm CNT}^{\rm opt}$
from Eqs.~\eqref{eq:scnt} and~\eqref{eq:muopt}, as follows
\begin{equation}
  S_{\rm CNT}^{\rm opt} = S_{\rm CNT}(\mu = \mu_{\rm opt}).
  \label{eq:sopt}
\end{equation}
In Figs.~\ref{fig:thermo-plot}(a-b), we show the optimum thermopower
values of many s-SWNTs with $0.5\le\diameter\le1.5\unitnm$ compared
with their corresponding band gaps as a function of diameter.  In
Fig.~\ref{fig:thermo-plot}(a), we plot the optimum thermopower
calculated from the BoltzTrap simulation (denoted by circles) and from
$S_{\rm CNT}^{\rm opt}$ (denoted by plus symbols) on the same scale.
We can see that the two methods show a good agreement.  From
Fig.~\ref{fig:thermo-plot}(a), the thermopower of s-SWNTs is also
found to increase as the tube diameter $\diameter$ decreases.  For
some s-SWNTs with $d_t < 0.6\unitnm$, such as those with $2n+m = 13$,
i.e. the $(5,3)$ and $(6,1)$ s-SWNTs, the thermopower can reach a
value more than $2000\unittep$.  These thermopower values are about
$6$--$10$ times larger than those found in common thermoelectric
materials~\cite{shakouri11-thermo,boukai08-silicon,heremans08-enhancement,pei11-thermo,poudel08-thermo}.

The larger thermopower for smaller-diameter s-SWNTs can be explained
by the relation of $S_{\rm CNT}$ with $E_g$ as shown in
Eq.~\eqref{eq:scnt} and by the fact that $E_g \propto
1/\diameter$~\cite{saito00-trig}.  The one-dimensional character of
the SWNT electronic DOS may also enhance the
thermopower~\cite{hicks93-thermo,hicks96-thermo}.  Here, we should
note that the thermopower of s-SWNTs as a function of diameter shows
the nanotube family pattern, in which the different SWNTs with the
same $2n+m$ can be connected and they make a clearly distinct branch
for $\mod(2n+m,3) = 1$ and $\mod(2n+m,3)=2$, known as the nanotube SI
and SII family branches, respectively~\cite{saito00-trig}. This
behavior is very similar to that found in the band gap as a function
of diameter shown in Fig.~\ref{fig:thermo-plot}(b), which is often
referred to as the Kataura
plot~\cite{saito00-trig,kataura99-opt,weisman03-opt}.  This result
also suggests that the measurement of the thermopower of a single
chirality s-SWNT sample might be able to predict an exact band gap
value of the s-SWNT.  In fact, the band gap is directly connected to
the thermopower as can be seen in the $S_\mathrm{CNT}$ formula
[Eq.~\eqref{eq:scnt}].

Finally, we would like to briefly discuss the issues of maximizing the
thermoelectric power factor, which is the numerator term in the $ZT$
formula.  There are two main issues to which we have to pay attention.
First, we may argue that, for s-SWNTs as a thermoelectric material, it
might still be impossible to obtain a large $ZT$ or a usable device at
a low doping level despite the fact that the optimum thermopower
values are obtained near $\mu =0$.  The reason is that the electrical
conductivity $\sigma$ can be very small near $\mu = 0$.  This fact is
also reflected in the conductivity equation as a function of $\mu$
[see Appendix B, Eq.~\eqref{eq:S13}].  However, compared to the bulk
materials, the one-dimensional materials such as s-SWNTs have smaller
effective mass $m^*$, which may enhance the electrical conductivity
due to the relation of $\sigma \propto (m^*)^{-1/2}$, as can also be
seen in Eq.~\eqref{eq:S13}.  Second, we may worry that, as we go to
smaller diameter s-SWNTs (in which the thermopower is optimized), the
electrical conductivity will instead be too small to maximize the
power factor.  However, we note that there is also a chirality
dependence which could enhance the electrical conductivity through the
effective mass relation.  As mentioned before, a smaller $m^*$ will
give a larger $\sigma$, and thus s-SWNTs which have both small
diameters and small $m^*$ might be useful as a thermoelectric material
even at relatively low doping levels.

\section{Summary}
\label{sec:summary}

We have shown the theoretically predicted behavior of the thermopower
of many s-SWNTs within a diameter range of $0.5$--$1.5\unitnm$.  We
derive a simple formula to calculate the thermopower of s-SWNTs from
their band gap, which enables us to predict the optimum thermopower
values.  The optimum thermopower value of an individual s-SWNT (p-type
or n-type) can be larger than $2000\unittep$ at room temperature for
diameters less than $0.6\unitnm$, such as the $(5,3)$ and $(6,1)$
s-SWNT.  Our results highlight potential properties of small diameter
s-SWNTs as a one-dimensional thermoelectric material with a giant
thermopower.  With the recent advances in the fabrication methods for
specific small diameter
s-SWNTs~\cite{liu11-gel,liu13-gel,nakai14-giantS}, we expect that the
further potential development of s-SWNT thermoelectric devices could
be realized in the near future.

\section*{Acknowledgments}
N.T.H. and A.R.T.N acknowledge the support from the Tohoku University
Program for Leading Graduate Schools. R.S. acknowledges MEXT Grants
Nos. 25107005 and 25286005. M.S.D acknowledges support from NSF grant
DMR-1004147.

\appendix

\section{Thermopower of nondegenerate semiconductors}
Here we derive a general formula for the thermopower of nondegenerate
semiconductors as a starting point before deriving the analytical
formula of $S_\mathrm{CNT}$ [Eq.~\eqref{eq:scnt}].  In the calculation
of the thermopower, we assume that the single wall carbon nanotubes
(s-SWNTs) are nondegenerate semiconductors.  The thermopower or the
Seebeck coefficient for a nondegenerate semiconductor can be
calculated by solving the Boltzmann transport equation under the
relaxation time approximation, which leads to the following
expression~\cite{goldsmid10-thermo}:
\begin{equation}
\label{eq:S1}
S=\frac{1}{qT}\frac{\displaystyle \int{\upsilon(\varepsilon) \tau(\varepsilon)
    \upsilon(\varepsilon) \dfrac{\partial f_0(\varepsilon)}{\partial
      \varepsilon} [\varepsilon-\mu]g(\varepsilon)
    d\varepsilon}}{\displaystyle \int{\upsilon(\varepsilon) \tau(\varepsilon)
    \upsilon(\varepsilon) \dfrac{\partial f_0(\varepsilon)}{\partial
      \varepsilon}g(\varepsilon) d\varepsilon}},
\end{equation}
where $q$, $\varepsilon$, $T$, and $\mu$ are the unit carrier charge,
electronic band energy, temperature, and chemical potential,
respectively. The variables $\upsilon(\varepsilon)$,
$\tau(\varepsilon)$, $f_0(\varepsilon)$, and $g(\varepsilon)$ are the
band carrier velocity, carrier relaxation (scattering) time,
Fermi-Dirac distribution function, and the density of states (DOS) per
unit volume, respectively, defined by
\begin{equation}
\label{eq:S2}
\upsilon^2(\varepsilon)=\frac{2\varepsilon}{m^*d},
\end{equation}
\begin{equation}
\label{eq:S3}
\tau(\varepsilon)=\tau_0\varepsilon^r,
\end{equation}
\begin{equation}
\label{eq:S4}
f_0(\varepsilon)=\frac{1}{1+e^{(\varepsilon-\mu)/{k_BT}}},
\end{equation}
\begin{equation}
\label{eq:S5}
g(\varepsilon)=\frac{1}{L^{3-d}2^{d-1}\pi^{d/2}\Gamma{\left(
    \frac{d}{2}\right)} }\left( \frac{2m^*}{\hbar^2}\right)^{d/2}
\varepsilon^{d/2-1},
\end{equation}
where $d=1,2,3$ denotes the dimension of the material, $m^*$ is the
effective mass of electrons or holes, $r$ is a characteristic
exponent, $\tau_0$ is the relaxation time constant, and $L$ is the
confinement length for a particular material dimension.  Substituting
Eqs.~\eqref{eq:S2}-\eqref{eq:S5} into Eq.~\eqref{eq:S1} yields
\begin{equation}
\label{eq:S6}
S=\frac{1}{qT}
\left(\mu-\dfrac{\displaystyle \int\varepsilon^{d/2+r+1}\dfrac{\partial
      f_0(\varepsilon)}{\partial \varepsilon}
    d\varepsilon}{\displaystyle \int\varepsilon^{d/2+r}\dfrac{\partial
      f_0(\varepsilon)}{\partial \varepsilon} d\varepsilon} \right).
\end{equation}

To simplify Eq.~\eqref{eq:S6}, we define the following variables: the
reduced band energy $\xi=\varepsilon/(k_B T)$, the reduced chemical
potential $\eta=\mu/(k_B T)$, and the Fermi-Dirac integral
$F_j(\eta)=\int\xi^j f_0(\xi) d\xi$.  Inserting these quantities into
Eq.~\eqref{eq:S6} gives
\begin{equation}
\label{eq:S7}
S=-\frac{k_{B}}{q}\left(\eta-\dfrac{\frac{d}{2}+r+1}{\frac{d}{2}+r}
 \times \dfrac{F_{d/2+r}}{F_{d/2+r-1}}\right).
\end{equation}
Since $(\xi-\eta)>3$ for nondegenerate semiconductors, we can use an
approximation of $F_j(\eta)\approx e^{\eta}\Gamma(j+1)$, where
$\Gamma(j)$ is the gamma function, to obtain
\begin{equation}
\label{eq:S8}
S=-\frac{k_{B}}{q}\left(\eta-\dfrac{\frac{d}{2}+r+1}{\frac{d}{2}+r}
\times \dfrac{\Gamma(\frac{d}{2}+r+1)}{\Gamma(\frac{d}{2}+r)}\right).
\end{equation}
Using the recursion formula $\Gamma(j+1)=j\Gamma(j)$, the thermopower
of nondegenerate semiconductors within the one-band model can be
written as
\begin{equation}
\label{eq:S9}
S=-\frac{k_{B}}{q}\left(\eta-\dfrac{d}{2}-r-1\right).
\end{equation}
This last equation is still insufficient to derive the thermopower of
s-SWNTs since the s-SWNTs are considered as nondegenerate
semiconductors with two energy bands.  In this case, we also need an
expression of electrical conductivity because the semiconductor within
the two-band model includes a conduction band for electrons and a
valence band for holes following the formula $S = (\sigma_n S_n +
\sigma_p S_p) / (\sigma_n + \sigma_p)$~\cite{goldsmid99-thermo}, where
$S_{n,p}$ and $\sigma_{n,p}$ are, respectively, the thermopower and
electrical conductivity for the n-type or p-type semiconductors.  The
expression specifying the electrical conductivity for a single energy
band is derived in Appendix~\ref{sec:elec}.

\section{Electrical conductivity for nondegenerate
semiconductors}
\label{sec:elec}

The electrical conductivity is expressed as~\cite{goldsmid10-thermo}
\begin{equation}
\label{eq:S10}
\sigma=-q^2\int\upsilon(\varepsilon) \tau(\varepsilon)
\upsilon(\varepsilon) \dfrac{\partial f_0(\varepsilon)}{\partial
  \varepsilon}g(\varepsilon) d\varepsilon.
\end{equation}
Substituting Eqs.~\eqref{eq:S2}-\eqref{eq:S5} and the Fermi-Dirac
integrals into Eq.~\eqref{eq:S10} yields
\begin{align}
\label{eq:S11}
\sigma = &\frac{2 q^2 \tau_0 \left(\frac{d}{2}+r\right)}{m^*d}
\frac{1}{L^{3-d}2^{d-1}\pi^{d/2}\Gamma(\frac{d}{2})}
\left(\frac{2m^*}{\hbar^2}\right)^{d/2} \notag\\
&\times (k_BT)^{d/2+r}F_{d/2+r-1}.
\end{align}
By applying the approximation $F_j(\eta)\approx e^{\eta}\Gamma(j+1)$
for nondegenerate semiconductors, we can write the electrical
conductivity,
\begin{align}
\label{eq:S12}
\sigma = &\frac{2 q^2 \tau_0 \left(\frac{d}{2}+r\right)}{m^*d}
\frac{1}{L^{3-d}2^{d-1}\pi^{d/2}\Gamma(\frac{d}{2})}
\left(\frac{2m^*}{\hbar^2}\right)^{d/2} \notag\\
&\times (k_BT)^{d/2+r}e^\eta\Gamma\left(\frac{d}{2}+r\right),
\end{align}
which finally becomes
\begin{equation}
\label{eq:S13}
\sigma =\frac{2 q^2 \tau_0 \left(\frac{d}{2}+r\right)
  (k_BT)^{d/2+r}\Gamma(\frac{d}{2}+r)}{d\ L^{3-d}2^{d/2-1}\pi^{d/2}\hbar^d
  \Gamma(\frac{d}{2})}(m^*)^{d/2-1}e^\eta.
\end{equation}
We will use Eq.~\eqref{eq:S13} for calculating the electrical conductivity to
derive the thermopower of two-band semiconductors in the next section.

\section{Thermopower of two-band semiconductors}
The thermopower of two-band semiconductors is defined
by~\cite{goldsmid99-thermo}
\begin{equation}
\label{eq:S14}
S=\dfrac{\sigma_n S_n + \sigma_p S_p}{\sigma_n + \sigma_p},
\end{equation}
where $\sigma_{n,p}$ and $S_{n,p}$ are expressed as
\begin{align}
\label{eq:S15}
\sigma_{n,p} =&\frac{2 q^2 \tau_0 \left(\frac{d}{2}+r\right)
(k_BT)^{d/2+r}\Gamma(\frac{d}{2}+r)}{d\ L^{3-d}2^{d/2-1}\pi^{d/2}\hbar^d
\Gamma(\frac{d}{2})} \notag\\
&\times (m_{n,p}^*)^{d/2-1}e^{\eta_{n,p}},
\end{align}
and
\begin{equation}
\label{eq:S16}
S_{n,p}=\mp\frac{k_{B}}{e}\left(\eta_{n,p}- \frac{d}{2}-r-1\right),
\end{equation}
respectively. Substituting Eqs.~\eqref{eq:S15} and \eqref{eq:S16} into
Eq.~\eqref{eq:S14}, and after doing some algebra, we can obtain
\begin{equation}
\label{eq:S17}
S=\frac{k_{B}}{e}\frac{\displaystyle \left(\eta_n-
  \frac{d}{2}-r-1\right)\frac{\sigma_n}{\sigma_p}-\left(\eta_p-
  \frac{d}{2}-r-1\right)}{\displaystyle\frac{\sigma_n}{\sigma_p}+1}.
\end{equation}
where ${\sigma_n}/{\sigma_p}=\left(
  {m_n^*}/{m_p^*}\right)^{d/2-1}e^{\eta_n-\eta_p}=Ae^{\eta_n-\eta_p}$,
with $A=\left( {m_n^*}/{m_p^*}\right)^{d/2-1}$. Here we have
$\eta_n-\eta_p=2\eta_{\mu}$ and $\eta_n+\eta_p=-\eta_{g}$, where
$\eta_{\mu} = \mu/(k_BT)$ and $\eta_{g}=E_g/(k_BT)$.  The thermopower
of nondegenerate semiconductors within the two-band approximation can
then be written in terms of $\eta_\mu$, $\eta_g$, $r$, $d$, and $A$ as
\begin{align}
\label{eq:S18}
S=\frac{k_{B}}{e}\left(\eta_\mu - \frac{\eta_g}{2}
-r-\frac{d}{2}-1 +\frac{\eta_g+2r+d+2}{Ae^{2\eta_\mu}+1}\right).
\end{align}

\begin{figure}[t]
  \centering
  \includegraphics[clip,width=85mm]{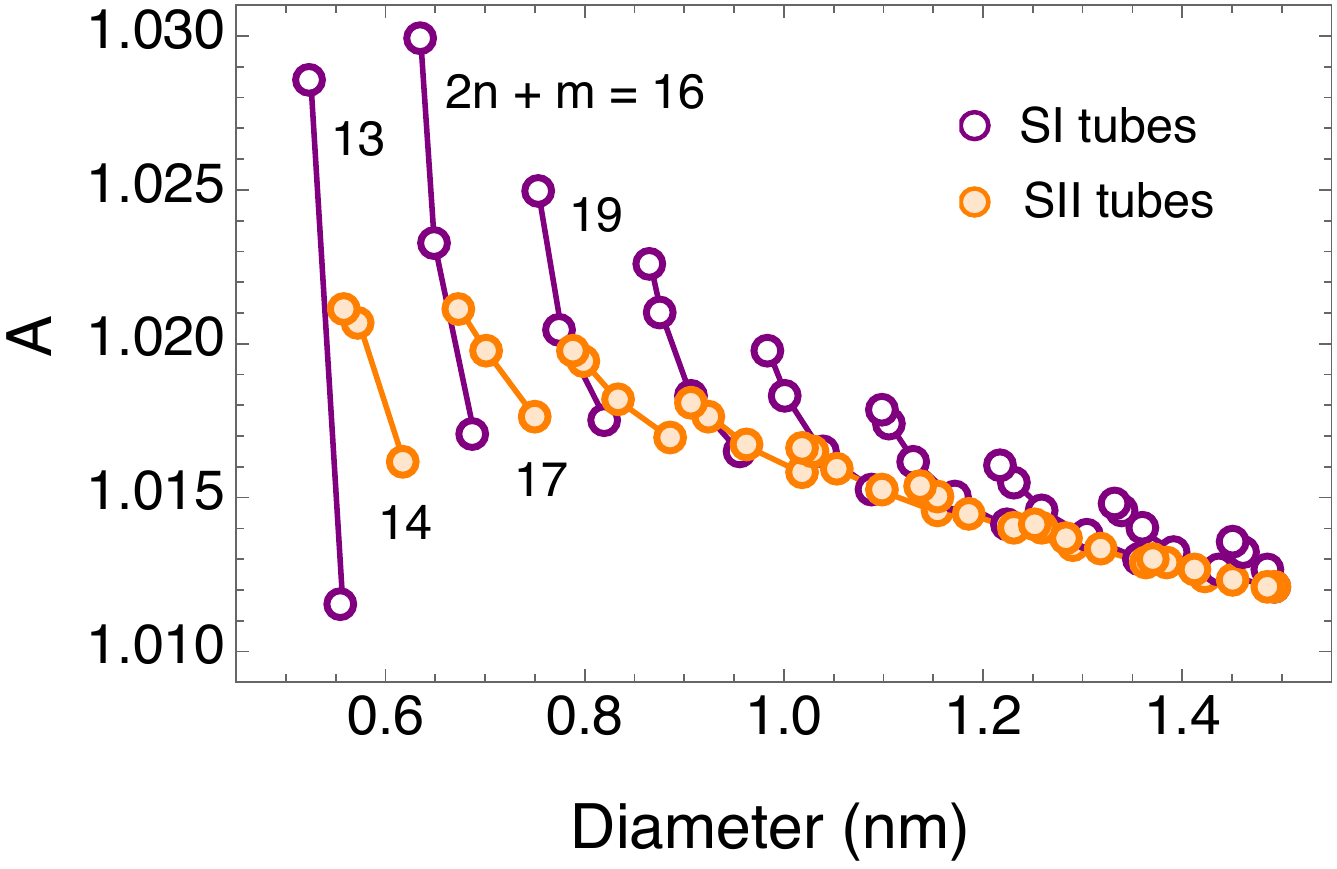}
  \caption{\label{fig:S1} $A=\left( {m_n^*}/{m_p^*}\right)^{-1/2}$ for
    s-SWNTs plotted as a function of the SWNT diameter.  SI and SII
    tubes correspond to the SWNTs having $\mod(2n+m,3) = 1$ and $2$,
    respectively.  Solid lines connect SWNTs with the same $2n+m$
    value.}
\end{figure}

\section{Thermopower of s-SWNTs}
We now finally have all the information needed to derive
$S_\mathrm{CNT}$.  Since s-SWNTs are one-dimensional, we have $d=1$
and $A=\left( {m_n^*}/{m_p^*}\right)^{-1/2}$. The electron and hole
effective masses $m_{n,p}^*$ in the s-SWNTs can be calculated using
the effective mass formula $m^* = \hbar^2 (d^2\varepsilon/dk^2)^{-1}$,
where $\varepsilon(k)$ is the electronic energy dispersion within the
extended tight binding (ETB) model~\cite{georgii04-etb}.  We can
obtain $A$ as a function of diameter, as can be seen in
Fig.~\ref{fig:S1}, in which we show $A$ within a diameter range of
$0.5$--$1.5$ nm.  In this diameter range, we have $A \approx 1$.  With
such an approximation, and also assuming that the carrier relaxation
time is constant [which gives $r = 0$ according to Eq.~\eqref{eq:S3}],
the thermopower of s-SWNTs is then given by
\begin{equation}
\label{eq:S19}
S_\mathrm{CNT}=\frac{k_{B}}{e}\left(\eta_\mu -
\frac{\eta_g}{2}-\frac{3}{2}+\frac{\eta_g+3}{e^{2\eta_\mu}+1}\right).
\end{equation}
The thermopower can be rewritten in terms of $\mu$ and $E_g$ as
\begin{align}
\label{eq:S20}
S_\mathrm{CNT}=\frac{k_B}{e}\left(\frac{\mu}{k_B T} - \frac{E_g}{2 k_B
    T} - \frac{3}{2} +\frac{E_g/k_B T +3}{e^{2\mu/k_B T} +1}\right).
\end{align}
where $k_B$ is the Boltzmann constant and $E_g$ is taken from the ETB
calculation~\cite{georgii04-etb}.  We see that Eq.~\eqref{eq:S20} is
nothing but Eq.~\eqref{eq:scnt}.  In this derivation, the reason why
we put $r = 0$ is that the electron relaxation time $\tau$ in s-SWNTs
is determined mainly by the electron-phonon interaction with the TW
phonon mode (see the main text, Sec.~\ref{sec:model}), where the
relaxation time is taken to be independent of the electron
energy~\cite{jiang05-elph}.  Therefore, we can write $\tau = \tau_0$
or equivalently $r=0$.

\section{Comparison between numerical and
analytical methods}

\begin{figure}[t]
  \centering
  \includegraphics[clip,width=85mm]{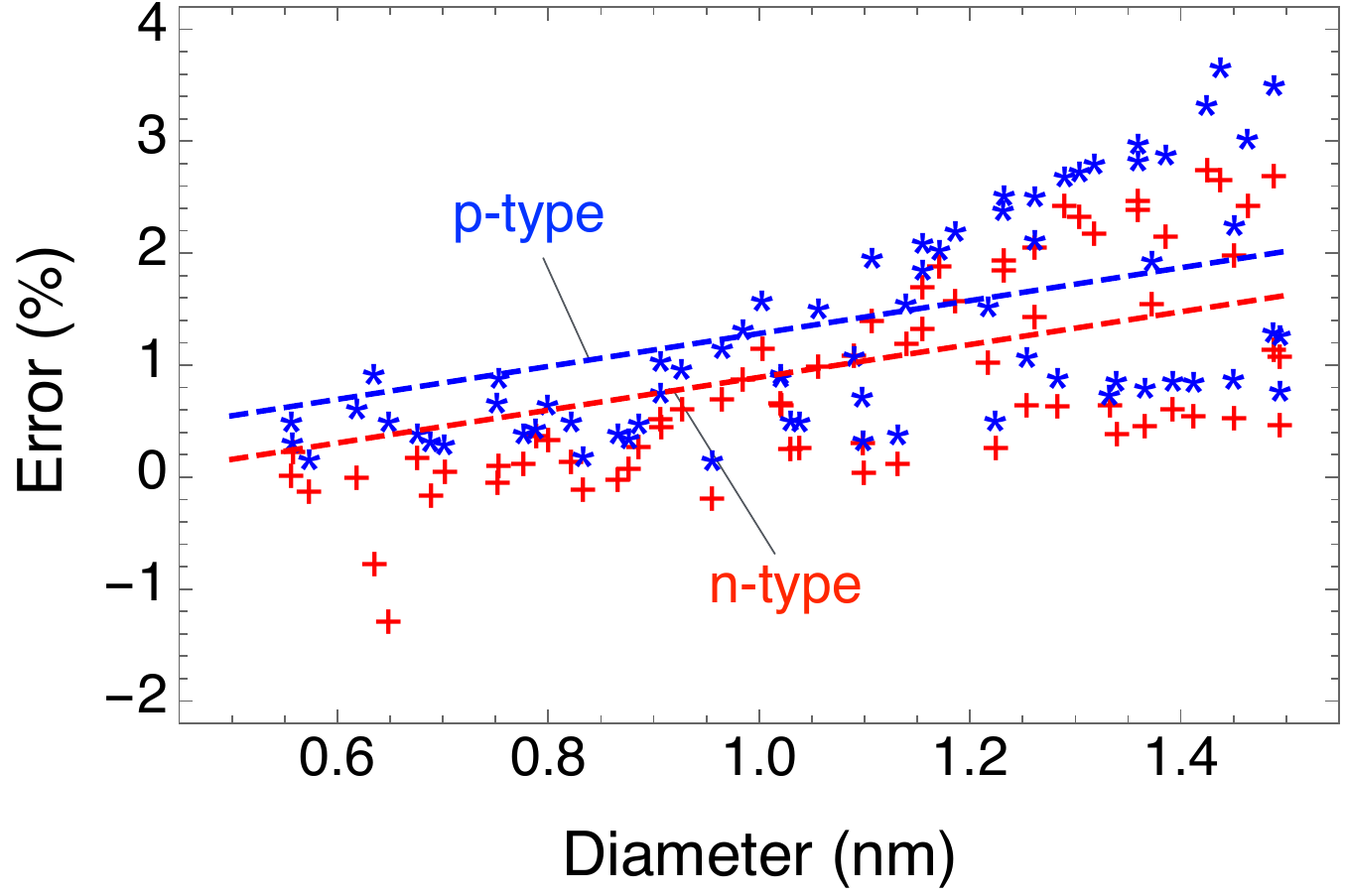}
  \caption{\label{fig:error} (Color online) The percentage error, or
    the discrepancy between the analytical and the numerical results
    of the thermopower calculations for each s-SWNT, is plotted versus
    the SWNT diameter.  The discrepancy increases linearly with
    increasing the SWNT diameter, as indicated by the fitted dashed
    lines.}
\end{figure}

To verify the accuracy of the $S_\mathrm{CNT}^\mathrm{opt}$ in fitting
the numerical results of the s-SWNT thermopower, we show in
Fig.~\ref{fig:error} the difference of the thermopower obtained from
the analytical and numerical calculations in terms of the error
percentage.  This error percentage variable is the difference in the
thermopower calculated by using the $S_\mathrm{CNT}^\mathrm{opt}$
formula with respect to the numerical results for each s-SWNT
diameter.  We obtain the error values ranging from $-2\%$ to $4\%$ for
both p-type and n-type s-SWNTs.  The error values increase with the
increase of the tube diameter because $E_g \propto 1/\diameter$ and
also because the formula for $S_{\rm CNT}$ [Eq.~\eqref{eq:scnt}] was
derived by assuming s-SWNTs as nondegenerate semiconductors.
Therefore, larger band gaps or smaller diameter s-SWNTs should be more
accurately fitted by our $S_{\rm CNT}$ approximation.

%\bibliographystyle{aip}
%\bibliography{nguyen}

\end{document}